\documentclass[runningheads]{lncse}

\usepackage[draft]{graphicx}

	 \def\be{\begin{equation}}
	 \def\ee{\end{equation}}
	 
	 \def\be{\begin{equation}}
	 \def\bea{\begin{eqnarray}}
	 \def\t{\tau}
	 \def\o{\over}
	 
	 \def\a{\alpha}
	 \def\b{\beta}
	 \def\e{\eta}
	 \def\ee{\end{equation}}
	 \def\eea{\end{eqnarray}}
	 \def\R{\rm {I\kern-.200em R}}
	 \def\C{\rm {I\kern-.520em C}}
	 \def\A{{\hat A}}
	 \def\D{{\hat D}}
	 \def\E{{\hat E}}
	 \def\at{\tilde {\alpha}}
	 \def\bt{\tilde {\beta}}

\begin{document}

\mainmatter

\title{An Exactly Solvable Two-Way Traffic Model with
Ordered Sequential Update}

\titlerunning{An Exactly Solvable Two-Way Traffic Model}

\author{
M. E. Fouladvand\inst{1,}\inst{3}\thanks{E-mail:
foolad@theory.ipm.ac.ir} 
\and 
H.-W. Lee\inst{2}
}

\authorrunning{M. E. Fouladvand and H.-W. Lee}

\institute{
Department of Physics, Sharif University of Technology, \\ 
P.O.Box 11365-9161, Tehran, Iran 
\and
School of Physics, Korea Institute for Advanced Study, \\
207-43 Cheongryangri-dong, Dongdaemun-gu, 
Seoul 130-012, Korea
\and
Institute for Studies in Theoretical Physics and Mathematics, \\
P.O.Box 19395-5531, Tehran, Iran
}

\maketitle

\begin{abstract}
Within the formalism of the matrix product ansatz, we study a two-species
asymmetric exclusion process with backward and forward site-ordered
sequential updates.  This model
describes a two-way traffic flow with a {\it dynamic impurity } and
shows a phase transition between the free flow and the traffic jam.
We investigate characteristics of this
jamming and examine similarities and differences between 
our results and those with the random sequential update~\cite{lee}. 
\end{abstract}

\section{Introduction}

A variety of phenomena can be modeled by 
the one dimensional asymmetric simple exclusion process (ASEP) and 
its generalizations (see \cite{book,schutzbook} and references therein). 
The model has a natural interpretation
as a description of traffic flow  
and constitutes a basis for more realistic ones \cite{nagel,wolf}. 
In traffic flow theories, the formation of traffic jams 
due to ``impurities'' is one of the fundamental problems. 

In the ASEP models, two kinds of impurities are discussed 
in the literature. The first one is ``dynamic'' impurities,
i.e., defective particles which jump with a rate  lower
than others~\cite{lee,derida2asep,mallick,evans}. In the 
traffic terminology, such moving defects can be visualized 
as slow cars on a road.
The other kind is ``static'' impurities 
such as imperfect links where the hopping rate is 
lower than in other links~\cite{janowsky,schutz1,emmerich,yukawa}. 
Both types of impurities can produce shocks 
in a system. 
Presently a limited amount of exact results is available 
for the shock formation and most of them are for models
with the random sequential update~\cite{lee,derida2asep,mallick}.

The implementation of the update is an essential part 
of the definition of a model and
it is of prime interest to determine whether distinct updating
schemes can produce different behaviors. 
The aim of this work is to investigate consequences of 
changing the updating scheme of the model. 
Here, we study an exactly solvable traffic model with two
types of ordered sequential updates, which is identical to
the model studied in Ref.~\cite{lee} except for the updating schemes.

\section{Model Definitions and Matrix Product Ansatz}

\subsection{Two-Way Traffic Model}

Consider two parallel one dimensional chains, 
each with $N$ sites and  the periodic boundary condition. 
There are $M$ cars and $K$ trucks in the first 
and the second chain, respectively, and
cars move to the right while trucks move to the left. 
 We introduce inter-chain interaction that forbids
a car and a truck to occupy two parallel sites
simultaneously. Then 
the state of the system can be described by
a single set of occupation numbers 
$(\t_1,\t_2,\cdots,\t_N)$
where $\t_i = 0$ (empty site), $1$ (occupied by a car), or 
$2$ (occupied by a truck).
There are
three possible hopping processes:
\begin{equation}
\begin{array}{lcl}
\mbox{(i) Car hopping with rate $\eta$} & \mbox{:} &
(1,0) \rightarrow (0,1) 
\\
\mbox{(ii) Truck hopping with rate $\eta\gamma$} & \mbox{:} &
(0,2) \rightarrow (2,0)
 \\
\mbox{(iii) Car-truck exchange with rate ${\eta\over \beta}$} & \mbox{:} &
(1,2) \rightarrow (2,1) \ .
\label{eq:dynamics}
\end{array}
\end{equation}
The reduction factor $\b$ ($1\le \b \le \infty$)
is related to the width of roads:
$\b =1$ corresponds to a very wide 
road or a highway with a lane divider, and $\b =\infty $ 
corresponds to a one lane road. 
Thus the value $1-1/\beta$ can be used as
a measure of the road narrowness. 

Recently the model is studied with
the random sequential update (RSU). 
In an infinitesimal time interval $dt$,
each link is updated with the probability
that is the product of the relevant rate and $dt$
[Since $\eta$ does not affect the dynamics at all 
in the RSU scheme, we choose $\eta=1$ for the RSU scheme].

Alternatively the ordered sequential update (OSU)
can be used.
One first chooses a particular site, i.e. the site $N$,  
and updates the state of the links
consecutively either in
the backward direction $(N,N-1), (N-1,N-2), \cdots, (1,N)$ 
[backward sequential update (BSU)] 
or in the forward direction
$(N,1), (1,2), \cdots, (N-1,N)$
[forward sequential update (FSU)].
In contrast to the RSU update, the time is discrete
in the BSU and FSU schemes, and
in each time step, hopping occurs in links with 
the probabilities that are equal to the rates
in Eq.~(\ref{eq:dynamics}).

\subsection{Matrix product state} 

The two-way traffic model is equivalent to 
a two-species ASEP and it can be 
solved exactly by the method of the matrix product state (MPS).
In the RSU scheme,  
the steady state weight $P_s$ of a given
configuration $(\t_1,\t_2,\cdots,\t_N)$ is proportional to 
the trace of the normal product of some matrices~\cite{lee}:
\be
P_s (\t_1,\t_2,\cdots,\t_N) \sim {\rm Tr}(X_1X_2\cdots X_N)
\label{MPSforRSU}
\ee
where $X_i=D$ for $\tau_i=1$; $E$ for $\tau_i=2$;
and $A$ for $\tau_i=0$, 
%
%
and these matrices satisfy the quadratic algebra 
\be
DE= D+E , \ \  \a AE=A , \ \  \b DA= A  \ \ (\a\equiv \b\gamma) \ .
\label{for1}
\ee

In a similar way, $P_s$ in the BSU (FSU) scheme can be written
as follows:
\be
P_s(\t_1,\t_2,\cdots,\t_N) \sim {\rm Tr}(X_1X_2\cdots {\hat X}_N) \ 
\label{for3}
\ee
where $X_i=D,E$, or $A$
($\hat{X}_N=\hat{D},\hat{E},$ or $\hat{A}$)
depending on $\t_i$ ($\t_N$).
Note that the matrices at the site $N$ are different from those
at other sites~\cite{raj98}. 
The presence of the hatted matrix in Eq.~(\ref{for3})
breaks the translational invariance of the problem.
In Ref.~\cite{Fouladvand}, it is shown that
with a simple assumption,
\be
\A=A+a, \ \ \D=D+d,\ \ \E=E +e \ ,
\label{for8}
\ee
where $a,d$, and $e$ are carefully chosen real numbers,
the relevant matrix algebra for $A,D,E$
becomes identical to
the one in Eq.~(\ref{for1}) except for
the renormalizations of $\alpha$ and $\beta$
to $\tilde{\alpha}$ and $\tilde{\beta}$.
In the BSU scheme, the choice 
$a=0, d=-\eta/\beta, e=\eta/(\beta-\eta)$ gives
$\at=\a( \b - \e) / (\b -\a \e)$,  
$\bt =\b$,
and in the FSU scheme, the choice
$a=0, d=\eta/(\beta-\eta), e=-\eta/\beta$ results in
$\at=\a$,  
$\bt =(\b - \e)/( 1-\e)$.

\section{Average Velocities}


In this section  we consider the special case where there is only one
truck and evaluate average velocities. 
The MPS method allows 
exact evaluations of velocities.
We first present the car velocity in the thermodynamic limit. 
In the BSU scheme,
\bea
\langle v_{\rm car}\rangle=  \ \ \left\{  
\begin{array}{ll} \displaystyle      
{\e \over 1-\e n}(1-n) 
  & {\rm if } \ \ \ n\bt \leq 1 \\ 
\displaystyle \rule{0mm}{8mm}
  {\e \over \b -\e}{1-n \over n}
  & {\rm if} \ \ \  n\bt \geq 1  \ ,
\end{array} \right.  
\label{v_car_BSU}
\eea
and in the FSU scheme,
\bea
\langle v_{\rm car}\rangle =  \ \ \left\{  \begin{array}{ll}      
\displaystyle 
  {\e (1-n)\o {1-\e (1-n)}} 
  & {\rm if } \ \ \ n\bt \leq 1 \\ 
\displaystyle  \rule{0mm}{9mm}
  {\e \over \b} {(1-n) \o n} & {\rm if } \ \ \ n\bt \geq 1 \ .
\end{array} \right.
\label{v_car_FSU}
\eea

\begin{figure}
\begin{minipage}{\textwidth}
\includegraphics[height=.5\textwidth,width=.5\textwidth]{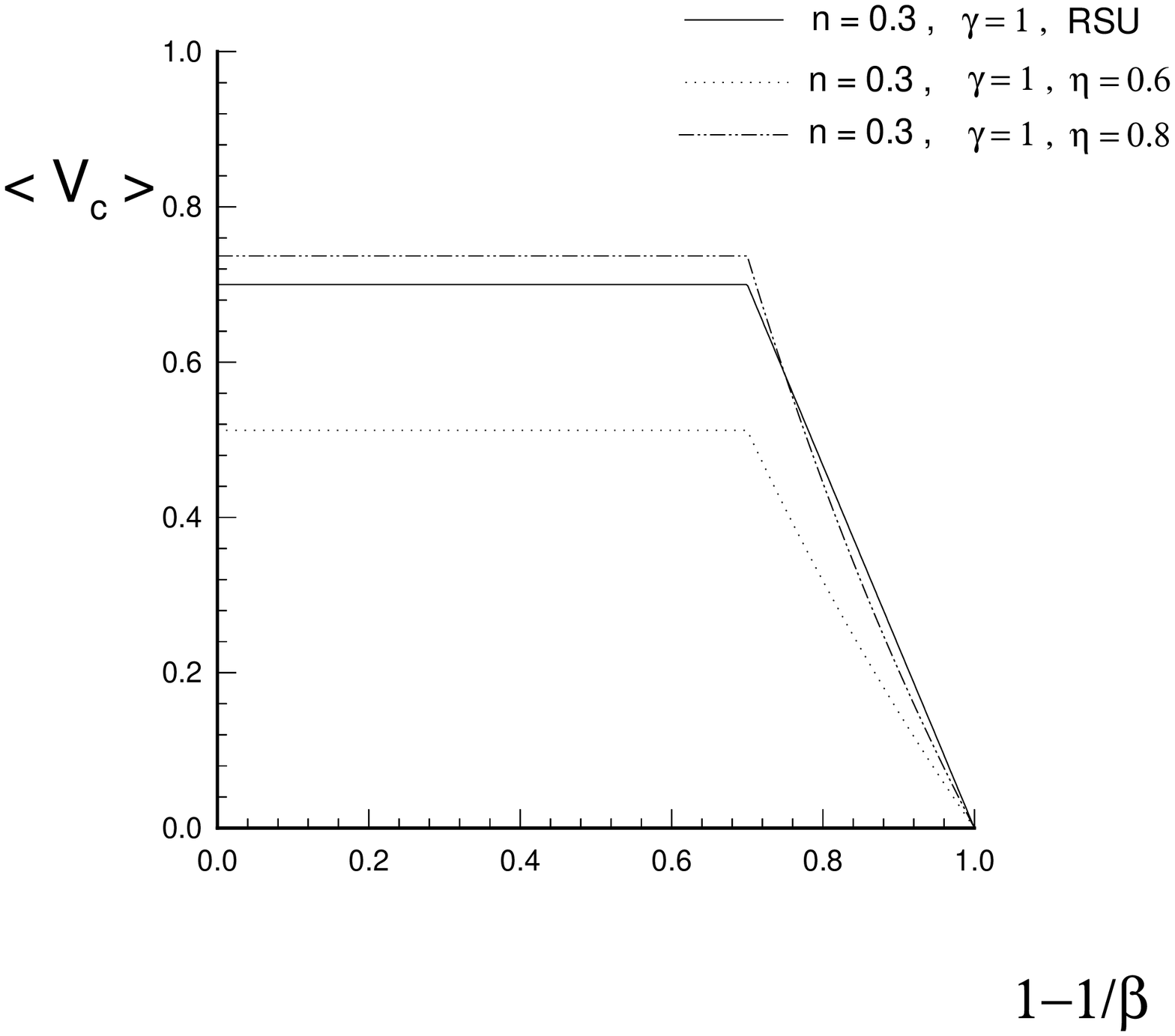}%
\includegraphics[height=.5\textwidth,width=.5\textwidth]{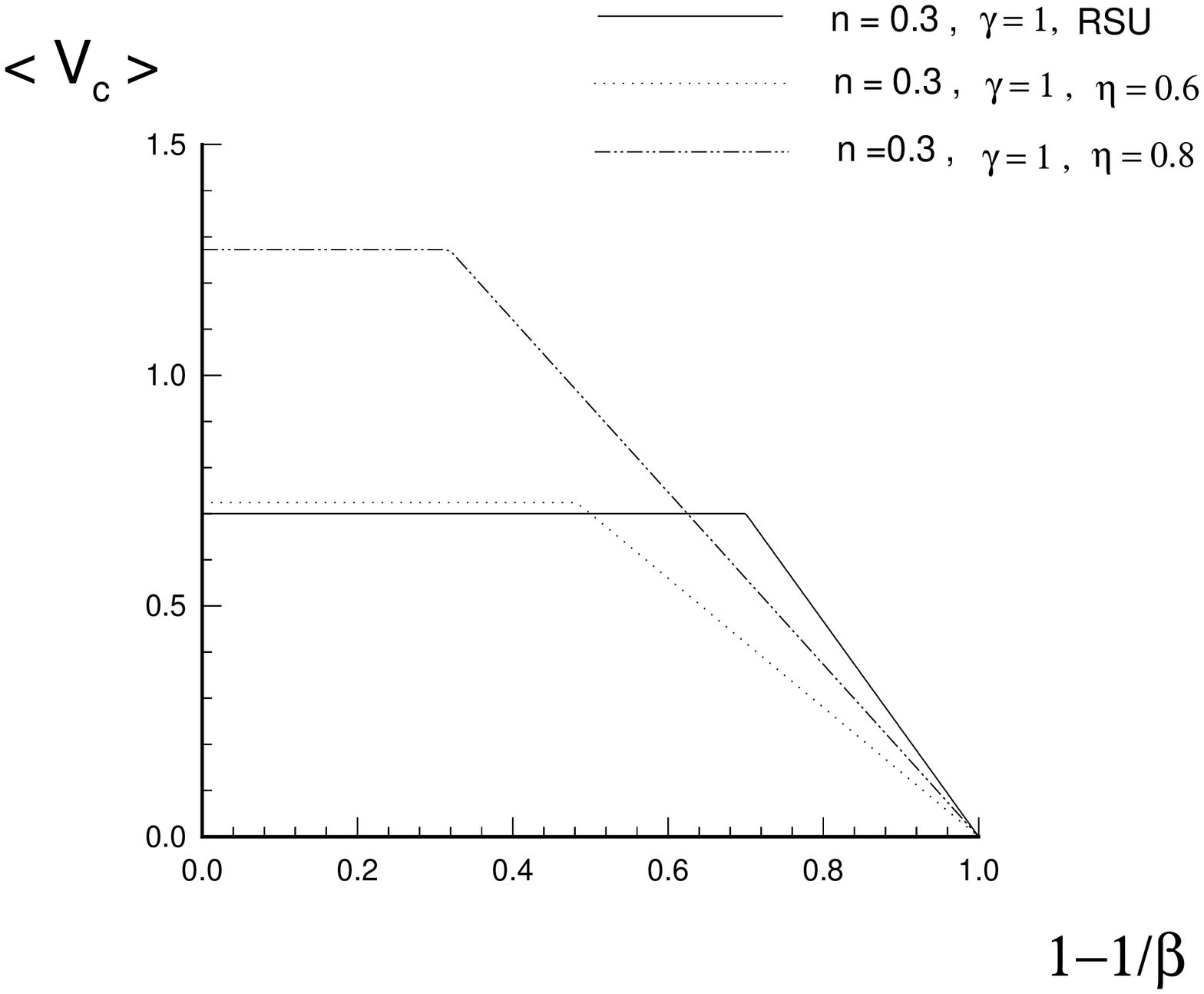}\\[0.2cm]
\mbox{}\hfill(a)\hfill\hfill(b)\hfill\mbox{}\\[0.4cm]
\includegraphics[height=.5\textwidth,width=.5\textwidth]{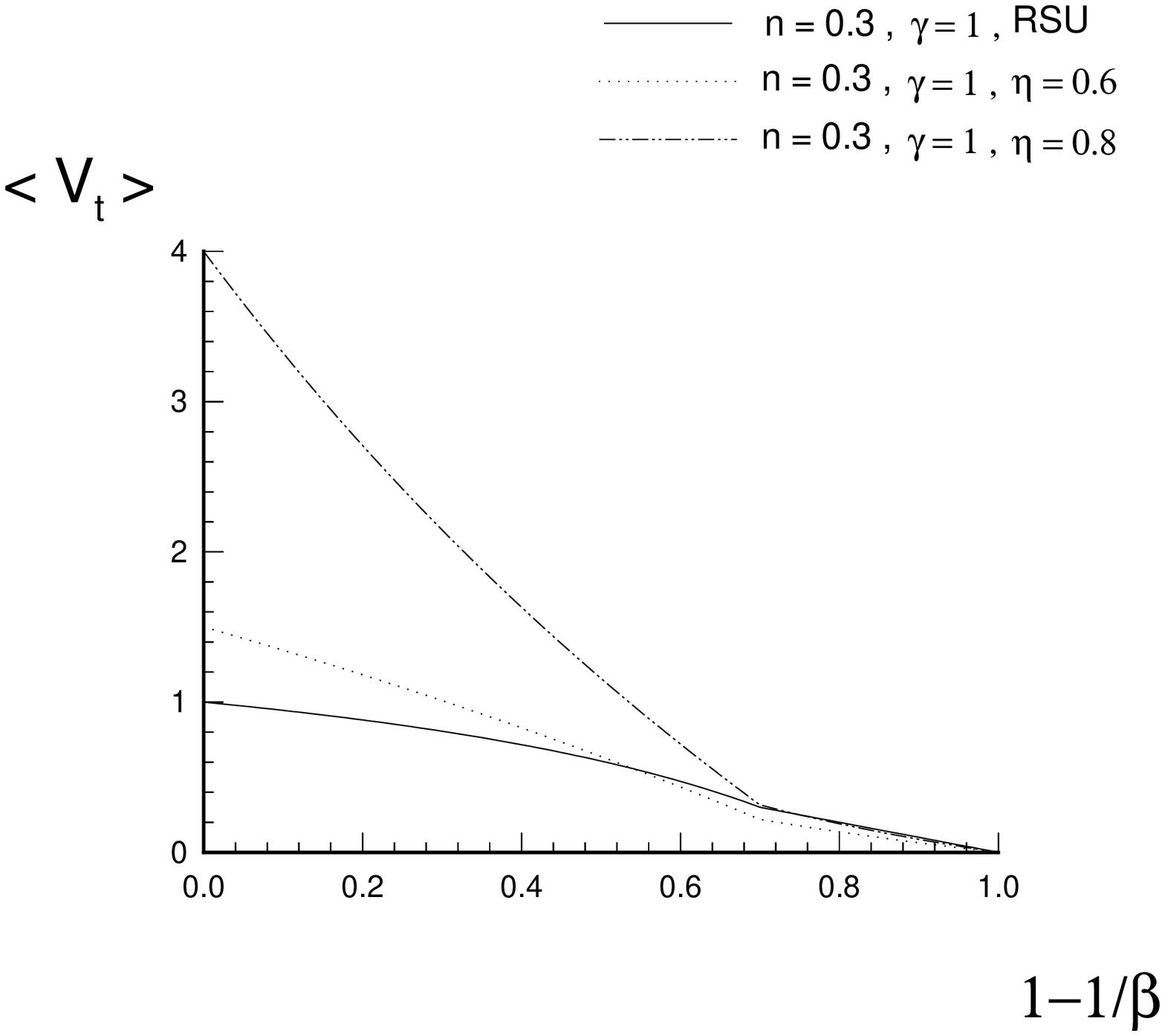}%
\includegraphics[height=.5\textwidth,width=.5\textwidth]{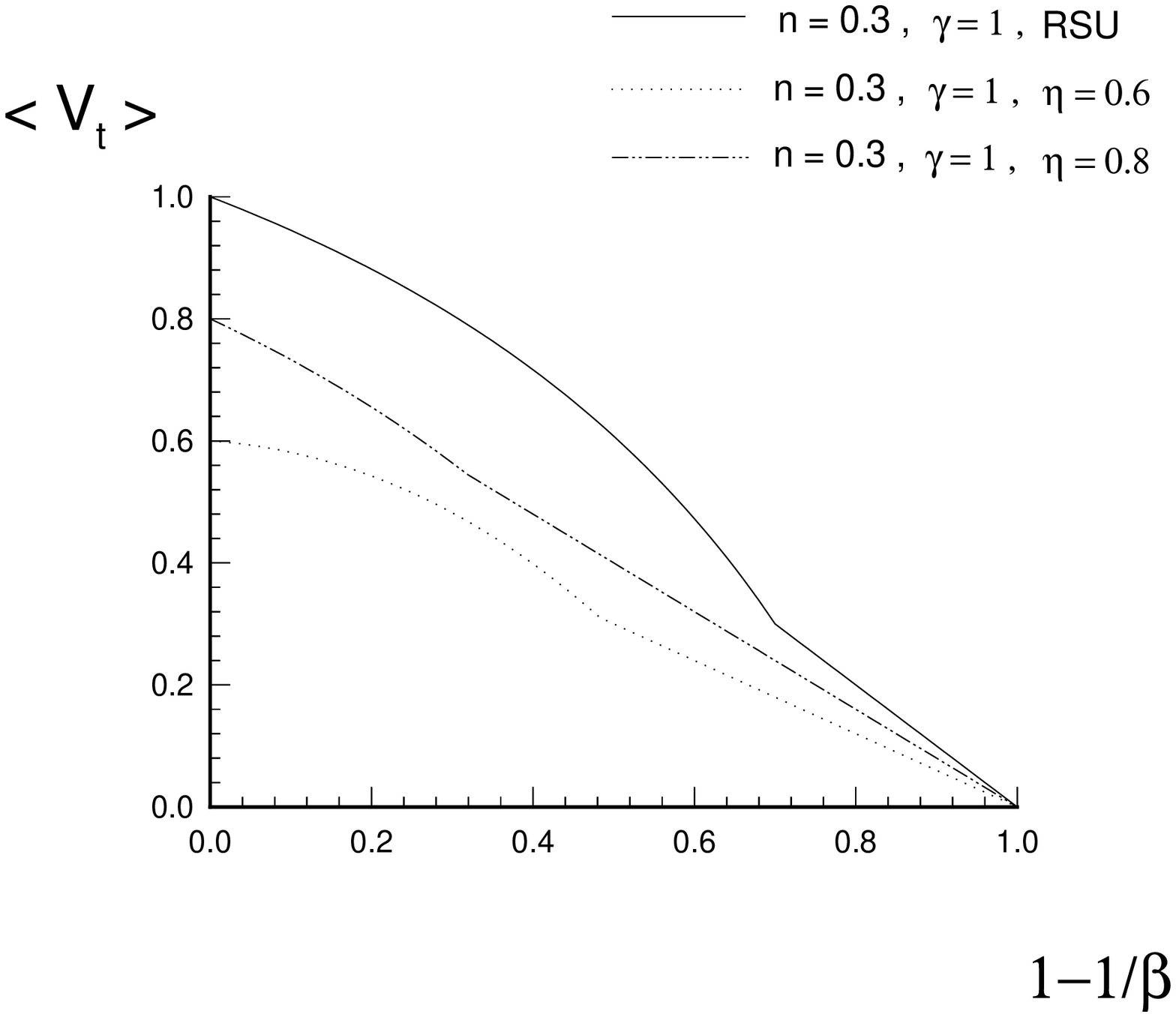}\\[0.2cm]
\mbox{}\hfill(c)\hfill\hfill(d)\hfill\mbox{}\\[0.4cm]
\end{minipage}
\caption{(a) Average velocity of cars
  in the BSU and RSU schemes for $n=0.3$
and different values of $\eta$.
(b) Average velocity of cars in the FSU and RSU schemes for $n=0.3$
and different values of $\eta$.
(c) Average velocity of the truck 
 in the BSU and RSU schemes for
$n=0.3$ and different values of $\eta$.
(d) Average velocity of the truck in the FSU and RSU schemes for
$n=0.3$ and different values of $\eta$.
After Ref.~\cite{Fouladvand}.
}
\end{figure}

Figs. 1(a) and (b) show the behaviors of $\langle v_{\rm car}\rangle$ 
in the two updating schemes as a function of 
the road narrowness $1-{1 \o \b}$ 
while the values of $\e , \gamma(={\a \over \b }) $ and $n$ are fixed.
In both schemes, a continuous phase transition is evident:
While $\langle v_{\rm car} \rangle$ is constant
below the critical narrowness,  
it begins to drop suddenly at the critical narrowness 
($n\tilde{\beta}=1$), generating a cusp. 
Thus above the critical narrowness, a single truck 
results in global effects.  
The functional dependence of the decrease differs
in the two updating schemes. 

The exact expressions for the truck velocity, on the other hand,
are rather lengthy especially below the critical
narrowness and we present the functional
dependence on the road narrowness only through 
figures [Figs. 1(c) and (d)]. For exact expressions,
see Ref.~\cite{Fouladvand}.
The appearance of a continuous transition
at the critical narrowness ($n\tilde{\beta}=1$) 
is clear and the functional dependence of the truck velocity
again differs in the two updating schemes.

\section{Density Profile}


In the RSU scheme, the probability to find a car at a site 
depends only on its relative distance $x$ to the truck
due to the translational invariance.
In Ref.~\cite{lee}, it is found that
in the high density  phase $n\b\geq 1$, 
the probability or density profile 
$\langle n(x)\rangle$ becomes 
\bea
\langle n(x)\rangle = \ \ \left\{  \begin{array}{cl} 
\displaystyle 1  & \mbox{for } {x \o N} 
  \le 
  {{n\b -1}\o {\b -1}}  \\ 
\displaystyle \rule{0mm}{8mm}{1\o {\b}}  & {\rm otherwise} \ .
\end{array} \right.
\label{nhighRSU}
\eea
Note that the system consists of two regions, 
a traffic jam region in front of the truck  
and a free flow region behind it. 
In the low density phase $n\b\le 1$, on the other hand,
the presence of the truck 
has only  local effects. In the thermodynamic limit,
the car density becomes 
\be
\langle n(x) \rangle = 
n \left \{ 1+ \frac{(\a + \b -1)(1-n)}{1-n+\a n}(n\b)^x 
  \right \} \ 
\label{nlowRSU}
\ee
which shows that the disturbance by the truck decays
exponentially with a characteristic length scale
$|\ln(n\b)|^{-1} \ .$


In the ordered sequential updates, complications
occur in the definition of the probability itself.
Since the choice of a particular site as a starting point
of the update breaks the translational invariance,
the probability to find a car at $x$ sites in front
of the truck depends not only on the relative distance
$x$ but also on the truck location.
This unnecessary complication can be avoided
by choosing the starting point in an even way.
In Ref.~\cite{Fouladvand}, it is found that with this ``homogeneous''
choice, the expressions for the probability 
in the BSU and the FSU scheme become
{\it identical} to Eqs.~(\ref{nhighRSU},\ref{nlowRSU})
except for the replacement of the bare parameters
$\alpha$ and $\beta$ with the renormalized ones
$\tilde{\alpha}$ and $\tilde{\beta}$.

\section{Concluding remarks}

Characteristics of an exactly solvable two-way
traffic model are investigated 
with the OSU scheme
and both qualitative and quantitative differences 
are found compared to those with
the RSU scheme \cite{lee}. 
Our approach is based
on the so-called matrix product formalism which allows analytic solutions. 
In the OSU schemes, the choice of a particular site as a starting
point of the update breaks the translational invariance of 
the steady state measure, which is also evident in the form 
of the MPS.  Thus an averaging over the different choices
of the update starting point is necessary to restore 
the translational invariance to the system. 
Performing the translationally invariant averaging,
it is found that 
for characteristics such as the density profile of cars and 
the density-density correlation function,
the differences in the updating schemes
can be fully taken into account by
the proper renormalization of the parameters $\a$ and $\b$.
However this is not the case with average velocities. 
Changing the update scheme affects velocities 
in a more complicated manner and the renormalization of
the parameters is not sufficient to account for
different behaviors of $\langle v_{\rm car } \rangle $ and 
$\langle v_{\rm truck } \rangle $ in different updating schemes.
Especially the dependence of $\langle v_{\rm car} \rangle$
and $\langle v_{\rm truck} \rangle$ on the road narrowness
$1-{1\over \b}$ can vary qualitatively depending 
on the updating schemes. 

\vspace{1cm} 
\noindent {\bf Acknowledgments} \\
 
M.E.F. would like to thank V. Karimipour for fruitful comments 
and D. Kim, R. Asgari for useful helps. H.-W.L. thanks D. Kim for
bringing his attention to this problem. 
H.-W.L. was supported by the Korea Science and Engineering Foundation
through the fellowship program and the SRC program at SNU-CTP.




\end{document}